\documentclass[12pt,a4paper]{article}

\usepackage{epsfig}
\usepackage{amsmath,amsfonts,amssymb}
\usepackage{t1enc}
\usepackage{cite}

\providecommand{\openone}{\leavevmode\hbox{\small1\kern-3.8pt\normalsize1}}

\newcommand{\Gp}{\Gamma_R}
\newcommand{\Gm}{\Gamma_L}
\newcommand{\Gz}{\Gamma_0}
\newcommand{\Gpm}{\Gamma_{R,L}}
\newcommand{\Fi}{F_i}
\newcommand{\Fpm}{F_{R,L}}
\newcommand{\fp}{F_R}
\newcommand{\fm}{F_L}
\newcommand{\fz}{F_0}
\newcommand{\rhp}{\rho_R}
\newcommand{\rhm}{\rho_L}
\newcommand{\rhpm}{\rho_{R,L}}

\newcommand{\thlw}{\theta_{\ell}^*}
\newcommand{\thlb}{\theta_{\ell b}}

\newcommand{\afb}{A_\mathrm{FB}}
\newcommand{\Ap}{A_+}
\newcommand{\Am}{A_-}
\newcommand{\Apm}{A_\pm}

\newcommand{\all}{A_{\ell \ell'}}

\newcommand{\allt}{\tilde A_{\ell \ell'}}

\newcommand{\alb}{A_{\ell b}}
\newcommand{\albt}{\tilde A_{\ell b}}
\newcommand{\vl}{V_L}
\newcommand{\vr}{V_R}
\newcommand{\gl}{g_L}
\newcommand{\gr}{g_R}

\begin{document}

\begin{flushright}
{\small
ATLAS scientific note \\
SN-ATLAS-2007-064 \\[2mm]
\today
}
\end{flushright}

\begin{center}
\begin{Large}
{\bf ATLAS sensitivity to $\boldsymbol{Wtb}$ anomalous couplings \\[2mm]
in top quark decays}
\end{Large}

\vspace{0.5cm}
J. A. Aguilar--Saavedra$^a$, J. Carvalho$^b$, N. Castro$^b$, A. Onofre$^b$,
F. Veloso$^b$  \\[0.2cm] 
{\it $^a$ Departamento de Física Teórica y del Cosmos and CAFPE, \\
Universidad de Granada, E-18071 Granada, Spain} \\[0.1cm]
{\it $^b$ LIP - Departamento de Física, \\
Universidade de Coimbra, 3004-516 Coimbra, Portugal}
\end{center}

\begin{abstract}
We study the sensitivity of the ATLAS experiment to $Wtb$ anomalous couplings
in top pair production with semileptonic decay,
$pp \to t \bar t \to W^+ b W^- \bar b$ with one of the $W$ bosons decaying 
leptonically and the other hadronically. Several observables are examined,
including the $W$ helicity fractions and new quantities recently introduced,
such as the ratios of helicity fractions and some angular asymmetries defined
in the $W$ rest frame. The dependence on anomalous couplings of all these
observables has been previously obtained. In this work we show that some of
the new observables also have smaller systematic uncertainties than the
helicity fractions, with a dependence on anomalous couplings similar
or stronger than for helicity fractions. Consequently, their measurement
can significantly improve the limits on anomalous couplings. Moreover, the
most sensitive measurements can be combined. In this case, the precision
achieved in the determination of $Wtb$ anomalous couplings can be of a few
percent in the semileptonic channel alone.
\end{abstract}

\section{Introduction}

The three generation structure of the standard model (SM) was completed with the
discovery of the top quark at Tevatron \cite{disc}. 
Its properties have already been directly investigated at colliders
\cite{Whel,tfcnc,tspin,tcross}
and, in particular, its mass has been
determined to a high accuracy \cite{topmass}, better than for any other
quark. However, the
determination of other fundamental properties, like spin and
couplings, requires larger top samples, which will be available at
LHC. In its first low luminosity phase (10 fb$^{-1}$/year)
LHC will produce several millions of top quarks per year and
experiment, mainly in pairs through gluon fusion $g g \to t \bar t$ and
quark-antiquark
annihilation $q \bar q \to t \bar t$, with a total cross section of
833 pb for a top mass $m_t = 175$ GeV \cite{topquark,Kidonakis:2003qe}.
Single top production \cite{single1,single2}
will also occur, dominated by the
process $b q \to t q'$, with an expected cross section of 306 pb
\cite{topquark}. Both processes will test the SM
predictions for the
fundamental properties of the top quark, and in particular they will allow us to
measure its couplings \cite{single1,prd,espriu,larios}. This fact is
specially important since, from a theoretical point of view, sizeable deviations
from the SM predictions for top couplings are possible in several SM
extensions, as for example in supersymmetry \cite{susy} and models of
dynamical symmetry breaking \cite{tech}. Indeed,
within the SM the $Wtb$ coupling is purely left-handed at the tree level,
and its size is given by the Cabibbo-Kobayashi-Maskawa
(CKM) matrix element $V_{tb} \simeq 1$. But in SM extensions
departures from the SM expectation for $V_{tb}$ are possible
\cite{largo,jose}, as well as new radiative
contributions to the $Wtb$ vertex.
These deviations might be observed in top pair and single top production
at LHC.

Top pair production takes place mainly through QCD interactions, thus
independently of the electroweak $Wtb$ coupling.
Additionally, it is likely that the top quark almost
exclusively decays in the channel $t \to W^+ b$.
Therefore, the cross section for $gg,q\bar q \to t \bar t
\to W^+ b W^- \bar b$ is insensitive to the size of the $Wtb$
vertex, as well as to its chiral and tensorial structure.
Still, the angular distributions of top decay products give
information about the $Wtb$ vertex structure
(up to a global multiplicative constant),
and thus they can be used to probe anomalous top couplings.\footnote{The
global normalisation of the $Wtb$ vertex can be determined in single top
production, whose cross section is proportional to $|V_{tb}|^2$ plus terms
involving anomalous couplings. Hence, the complete determination of the
$Wtb$ coupling requires the combination of measurements in single top
and top pair production.}
In the rest frame of a decaying top quark, the energies of the
$W$ boson and $b$ quark are fixed by the two-body kinematics. Therefore,
non-standard $Wtb$ interactions can only influence the following groups of
observables:
\begin{enumerate}
\item The total width $\Gamma(t \to Wb)$, which is very difficult
to measure at LHC.\footnote{The SM expectation,
$\Gamma(t \to Wb) \sim 1.6$ GeV at the tree level, is one order of
magnitude smaller than the width of the top invariant mass distribution
reconstructed in the detector, which is about 12 GeV (see for example
Ref.~\cite{Hubaut:943202}). Thus, deviations from the SM prediction for the
top quark width are not likely to be observable.}
\item The helicity fractions of the $W$ boson, which also determine the
angular distributions of its decay products in the $W$ rest frame
and their energy distributions in the top rest frame.
\item The angular distribution of the $W$ in the top rest frame, with respect
to the top spin direction.
\end{enumerate}
The second class of observables, those related to $W$ helicity fractions, may
be defined (and in principle measured) for the decay of a top quark
independently of the production mechanism, centre of mass energy, etc.
In particular, these observables can be measured in single top as well
as in top pair production.
Previous literature \cite{marsella} has already studied the sensitivity of
the ATLAS experiment \cite{ATLAS} for the measurement of helicity fractions
in top pair production and subsequent semileptonic or dileptonic decay.
Here we extend that analysis by including additional observables defined in
Ref.~\cite{paper1}: the ratios of helicity fractions (denoted as
``helicity ratios'') and some new angular symmetries defined in the $W$ rest
frame.
The dependence on anomalous couplings of the observables in the second
class has been obtained in Ref.~\cite{paper1}, including quadratic terms and
keeping the $b$ quark mass nonzero. In particular, it has been found that
some of the new observables in Ref.~\cite{paper1} have a stronger parametric
dependence on anomalous couplings than helicity fractions.
In this work we study in detail the ATLAS
sensitivity for their measurement in the semileptonic channel, paying a
special attention to systematic uncertainties, both the theoretical ones
and those related to the measurement in a real detector. 
We will eventually find that some of the new observables have smaller
systematic uncertainties, and that their measurement can significantly
improve the precision in the determination of anomalous couplings.

This paper is organised as follows. In section \ref{sec:2} we briefly set our
notation and define the observables studied. A more extensive discussion of
the theoretical aspects such as the relations among observables, their
analytical expressions and plots of their dependence on
anomalous couplings can be found in Ref.~\cite{paper1}.
In section \ref{sec:3} the generation of the $t \bar t$ signal and backgrounds
is outlined, together with the selection criteria used to analyse them.
In section \ref{sec:4} we present our results for the expected experimental
measurement of the observables
considered, and in section \ref{sec:5} we discuss their
implications for the experimental determination of $Wtb$ anomalous couplings.
Section \ref{sec:6} is devoted to our conclusions.

\section{The effective $\boldsymbol{Wtb}$ vertex and angular distributions in
$W$ rest frame}
\label{sec:2}

The most general $Wtb$ vertex containing terms up to dimension
five can be written as
\begin{eqnarray}
\mathcal{L} & = & - \frac{g}{\sqrt 2} \bar b \, \gamma^{\mu} \left( \vl
P_L + \vr P_R
\right) t\; W_\mu^- \nonumber \\
& & - \frac{g}{\sqrt 2} \bar b \, \frac{i \sigma^{\mu \nu} q_\nu}{M_W}
\left( \gl P_L + \gr P_R \right) t\; W_\mu^- + \mathrm{h.c.} \,,
\label{ec:1}
\end{eqnarray}
with $q=p_t-p_b$ the $W$ boson momentum. The new anomalous couplings $\vr$,
$\gl$ and $\gr$
\cite{prd,paper1} can be related to $f_1^R$, $f_2^L$ and $f_2^R$ in
Ref.~\cite{marsella} (and references therein) as $f_1^R=\vr$,
$f_2^L=-\gl$ and $f_2^R=-\gr$.  
If we assume CP is conserved these couplings can be taken to be real.
Within the SM, $\vl \equiv V_{tb} \simeq 1$ and the other couplings vanish
at the tree level, while small nonzero values are generated at one loop level
in the SM \cite{korner} and its extensions (see for example
Refs.~\cite{susy,tech}).

The measurement of angular distributions and asymmetries in top decays can
only determine ratios of couplings. (Besides, a moderate deviation from
$\vl \simeq 1$ is not visible in top pair production and decay, as long as
the top quark mainly decays to $W^+ b$ and all other channels are rare.)
Then, the value of $\vl$ sets the global scale for the measurement of
$\vr$, $\gl$ and $\gr$ in top decays. In this work we will normalise
$\vl$ to unity, and the limits on anomalous couplings
presented correspond to $\vl = 1$. For any other value, the 
corresponding limits on anomalous couplings can be obtained by
multiplying by the new $\vl$.

It must be noted that, apart from the direct measurement at LHC, low-energy
measurements already set indirect limits on
non-standard $Wtb$ couplings. 
The size of a $\vr$ term  is constrained by the measured rate of
$\mathrm{Br}(b \to s \gamma) = (3.3 \pm 0.4) \times 10^{-4}$ \cite{PDB}.
A right-handed coupling $|\vr| \gtrsim 0.04$ would in principle give a too
large
contribution to this decay \cite{vtbrbound} which, however, might be (partially)
cancelled with other new physics contributions. Hence, the bound $|\vr|
\leq 0.04$ is model
dependent and does not substitute a direct measurement of this coupling.
For $\gl$ the limits from $b \to s \gamma$ are of the same order, while for
$\gr$ they are much looser \cite{misiak}. Besides, if one allows all
anomalous couplings to be nonzero, direct and indirect
limits turn out to be complementary, because they constrain different
combinations of anomalous couplings.

As we have already pointed out,
the polarisation of the $W$ bosons produced in the top decay is sensitive
to non-standard $Wtb$ couplings \cite{kane}.
$W$ bosons can be
produced with positive, negative or zero helicity, with corresponding partial
widths $\Gp$, $\Gm$, $\Gz$ which depend on $\vl$, $\vr$, $\gl$ and $\gr$.
(General expressions for $\Gp$, $\Gm$, $\Gz$ in terms of these couplings
can be found in Ref.~\cite{paper1}.)
Their absolute measurement is rather difficult, so it is convenient to consider
instead the helicity fractions $\Fi \equiv \Gamma_i/\Gamma$, with
$\Gamma = \Gp + \Gm + \Gz$ the total width for $t \to Wb$.
Within the SM, $\fz = 0.703$, $\fm = 0.297$, $\fp = 3.6 \times 10^{-4}$
at the tree level, for $m_t = 175$~GeV, $M_W = 80.39$~GeV, $m_b = 4.8$~GeV. 
We note that $\fp$ vanishes in the $m_b=0$ limit
because the $b$ quarks produced in top decays have
left-handed chirality, and for vanishing $m_b$ the helicity and chirality
states coincide. 
These helicity
fractions can be measured in leptonic decays $W \to \ell \nu$. 
Let us denote by $\thlw$ the angle between the charged lepton
three-momentum in the $W$ rest frame and the $W$ momentum in the $t$ rest frame.
The normalised angular distribution of the charged lepton can be written as
\begin{equation}
\frac{1}{\Gamma} \frac{d \Gamma}{d \cos \thlw} = \frac{3}{8}
(1 + \cos \thlw)^2 \, \fp + \frac{3}{8} (1-\cos \thlw)^2 \, \fm
+ \frac{3}{4} \sin^2 \thlw \, \fz \,,
\label{ec:dist}
\end{equation}
with the three terms corresponding to the three helicity states and vanishing
interference \cite{dalitz}.  A fit to the
$\cos \thlw$ distribution allows to extract from experiment the values of $\Fi$,
which are not independent but satisfy $\fp + \fm + \fz = 1$.
From these measurements one can constrain the anomalous couplings in
Eq.~(\ref{ec:1}). Alternatively, from this
distribution one can measure the helicity ratios \cite{paper1}
\begin{equation}
\rhpm \equiv \frac{\Gpm}{\Gz} = \frac{\Fpm}{\fz} \,,
\label{ec:rho}
\end{equation}
which are independent quantities and take the tree-level values
$\rhp = 5.1 \times 10^{-4}$, $\rhm = 0.423$ in the SM.
As for the helicity fractions, the measurement of helicity ratios
sets bounds on $\vr$, $\gl$ and $\gr$.

A third and simpler method to extract information about the $Wtb$ vertex is
through angular asymmetries involving the angle $\thlw$.
For any fixed $z$ in the interval $[-1,1]$, one can define an asymmetry
\begin{equation}
A_z = \frac{N(\cos \thlw > z) - N(\cos \thlw < z)}{N(\cos \thlw > z) +
N(\cos \thlw < z)} \,.
\end{equation}
The most obvious choice is $z=0$, giving the forward-backward (FB) asymmetry
$\afb$ \cite{lampe,prd}.\footnote{Notice the difference in sign with respect
to the definitions in Refs.~\cite{lampe,prd}, where the angle $\thlb = \pi -
\thlw$ between the charged lepton and $b$ quark is used.}
The FB asymmetry is related to the $W$ helicity fractions by
\begin{equation}
\afb = \frac{3}{4} [\fp - \fm] \,.
\label{ec:afb}
\end{equation}
Other convenient choices are $z = \mp (2^{2/3}-1)$. Defining
$\beta = 2^{1/3}-1$, we have
\begin{eqnarray}
z = -(2^{2/3}-1) & \rightarrow & A_z = \Ap = 3 \beta [\fz+(1+\beta) \fp] \,,
\notag \\
z = (2^{2/3}-1) & \rightarrow & A_z = \Am = -3 \beta [\fz+(1+\beta) \fm] \,.
\label{ec:apm}
\end{eqnarray}
Thus, $A_+$ ($A_-$) only depend on $\fz$ and $\fp$ ($\fm$).
The SM tree-level values of these asymmetries are
$\afb = -0.2225$, $\Ap = 0.5482$, $\Am = -0.8397$.
They are very sensitive to anomalous $Wtb$ interactions, and their
measurement
allows us to probe this vertex without the need of a fit to the $\cos \thlw$
distribution. We also point out that with a measurement of two of these
asymmetries the helicity fractions and ratios can be reconstructed. For
instance, using Eqs.~(\ref{ec:apm}) and requiring $\fp + \fm + \fz = 1$, it is
found that
\begin{eqnarray}
\fp & = & \frac{1}{1-\beta} + \frac{\Am - \beta \Ap}{3 \beta(1-\beta^2)} \,,
 \notag \\
\fm & = & \frac{1}{1-\beta} - \frac{\Ap - \beta \Am}{3 \beta(1-\beta^2)} \,,
 \notag \\
\fz & = & - \frac{1+\beta}{1-\beta} + \frac{\Ap - \Am}{3 \beta (1-\beta)} \,.
\label{ec:inv}
\end{eqnarray}

\section{Simulation of signals and backgrounds and event selection}
\label{sec:3}

The $t \bar t \to W^+ b W^- \bar b$ events in which one of the $W$
bosons decays hadronically and the other one in the leptonic channel
$W \to \ell \nu_\ell$ (with $\ell=e^\pm,\mu^\pm$), are considered as signal
events. (From now on, the $W$ boson decaying hadronically and its parent
top quark will be named as ``hadronic'', and the $W$ decaying
leptonically and its parent top quark will be called ``leptonic''.)
Any other decay channel of the $t \bar t$ pair constitutes a background
to this signal. Top pair production, as well as the background from single top
production, is generated with TopReX 4.10~\cite{toprex} with default settings.
Further backgrounds
without top quarks in the final
state, i.e. $b\bar{b}$, $W+$jets, $Z/\gamma^*+$jets, $WW$, $ZZ$ and $ZW$
production processes, are generated using PYTHIA 6.206~\cite{pythia}.
In all cases we use CTEQ5L parton distribution functions (PDFs)~\cite{cteq}.
Events are hadronised using PYTHIA, taking also
into account initial state radiation (ISR), final state radiation (FSR)
and pile-up.

The generated background and signal events are passed through the ATLAS
fast simulation packages ATLFAST 2.53~\cite{atlfast} and
ATLFASTB~\cite{atlfast}. These packages
simulate the energy deposition in the calorimeter cells of all the
stable particles in each event.
The calorimeter cells are clustered within a cone of
$\Delta R = \sqrt{(\Delta\phi)^2 + (\Delta\eta)^2} = 0.4$, with $\phi$ the
azimuthal angle and $\eta$ the pseudorapidity. Cells with transverse energy
$E_T>1.5$~GeV are used as cluster seeds and the cone algorithm is applied
in decreasing order of $E_T$. Only clusters with $E_T>5$~GeV are
considered. The polar angle and the momentum of photons are smeared
according to Gaussian parameterisations. For electrons, their momenta are
smeared according to a Gaussian parameterisations. The momentum of each
muon is smeared according to a resolution which depends on the transverse
momentum $p_T$, as well as on $|\eta|$ and $\phi$.
The photon (electron) energy resolution is $\delta
E/E < 2.9\%$~ ($3.3\%$), for $E>20$~GeV. The transverse momentum
resolution of muons with $p_T<100$~GeV is $\delta p_T/p_T\lesssim
2\%$. Photons, electrons and muons are selected only if they have
$|\eta|<2.5$ and $p_T>5$~GeV ($p_T>6$~GeV for muons). They are
classified as isolated if the transverse energy of the cluster associated
to the particle, inside a cone of $\Delta R = 0.2$, does not exceed
the particle energy by $10$~GeV, and the $\Delta R$ from other energy
clusters must be above 0.4. The clusters of energy depositions not
associated to isolated photons, electrons or muons are used for the jet
reconstruction. Their momenta are smeared according to a Gaussian
distribution which depends on $|\eta|$. Jets are selected if they have
$E_T>10$~GeV. For $E>20$~GeV, the jet energy resolution is better than 12\%
(for pseudorapidities $|\eta| < 3$) and better than 24\% (for $|\eta| > 3$).
The missing transverse
momentum is estimated by summing the transverse momentum of the isolated
photons, electrons, muons and jets. The non-isolated muons and the
clusters of energy deposition which are not associated to isolated photons,
electrons, muons or jets, are also taken into account. In the ATLAS
detector, it will be possible to identify $b$ jets with $|\eta|<2.5$ by
using $b$ tagging tools. The algorithm was simulated by setting a
$b$-tagging efficiency to 60\%, with contamination factors set to 14.9\%
and 1.1\% for $c$ jets and light jets, respectively (the latter from light
quark, gluon and tau leptons). In order to check the
dependence of the analysis with the $b$-tagging efficiencies, different
values, 50\% and 70\% (corresponding to the expected $b$-tag variation
within the interesting signal transverse momentum range), were also
considered for the systematic studies, with contamination factors of 9.2\%
(0.4\%) and 23.3\% (2.9\%) for $c$-jets (light jets).

Due to the hadronisation and FSR, the jets are reconstructed
with less energies than those from the original quarks or gluons. The jets
energies are calibrated by the ATLFASTB package, by applying a
calibration factor,
$K^{\mathrm{jet}}=p_T^{\mathrm{parton}}/p_T^{\mathrm{jet}}$, which is the
ratio between the true parton energy and the reconstructed jet energy,
obtained from reference samples \cite{atlfast}. The
calibration factor depends on $p_T$ and is different for $b$-tagged
and light jets.

\begin{figure}[p]
\begin{center}
\begin{tabular}{ccccc}
\mbox{\epsfig{file=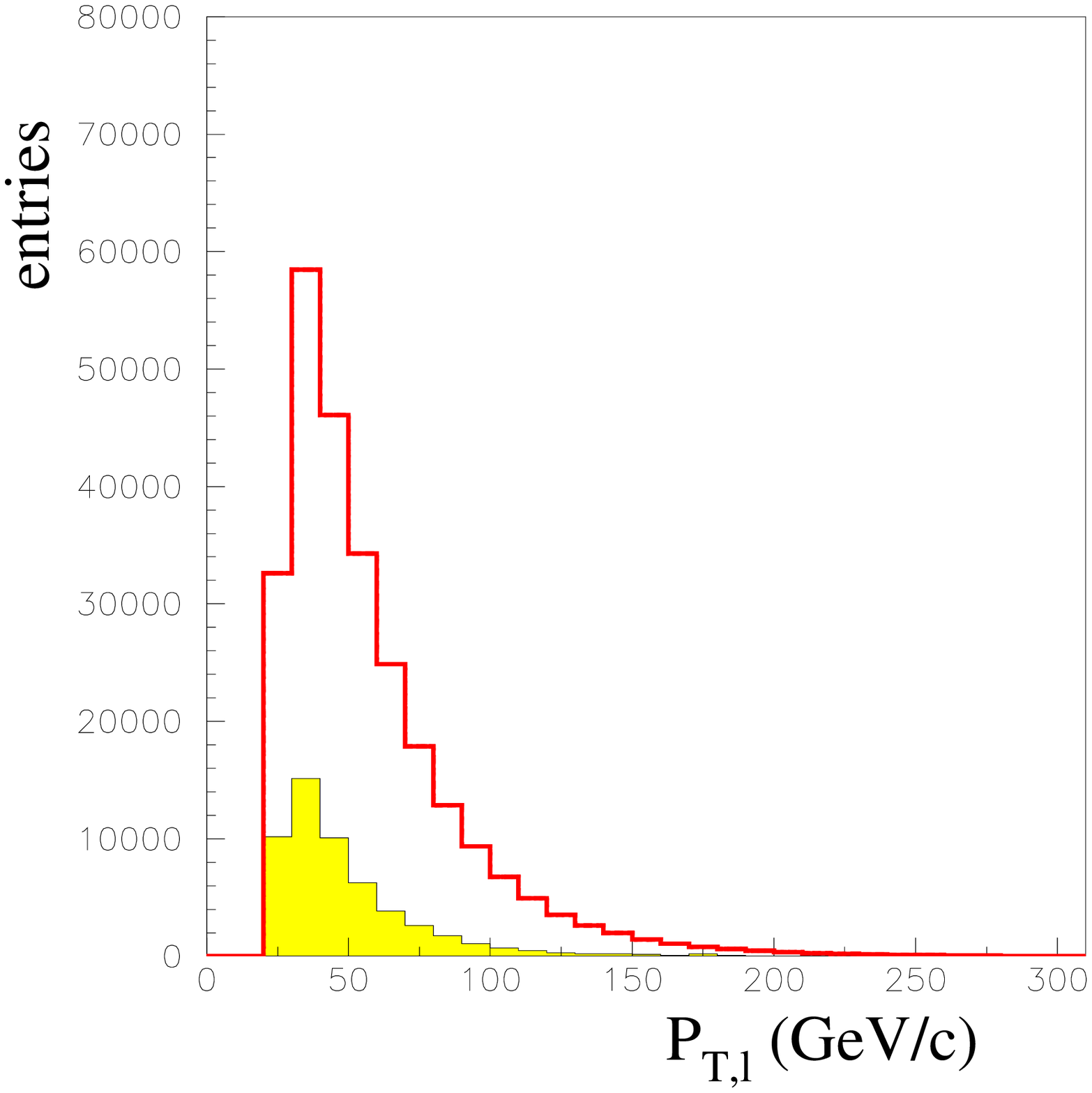,
height=5cm,clip=}} & \hspace{-6mm} &
\mbox{\epsfig{file=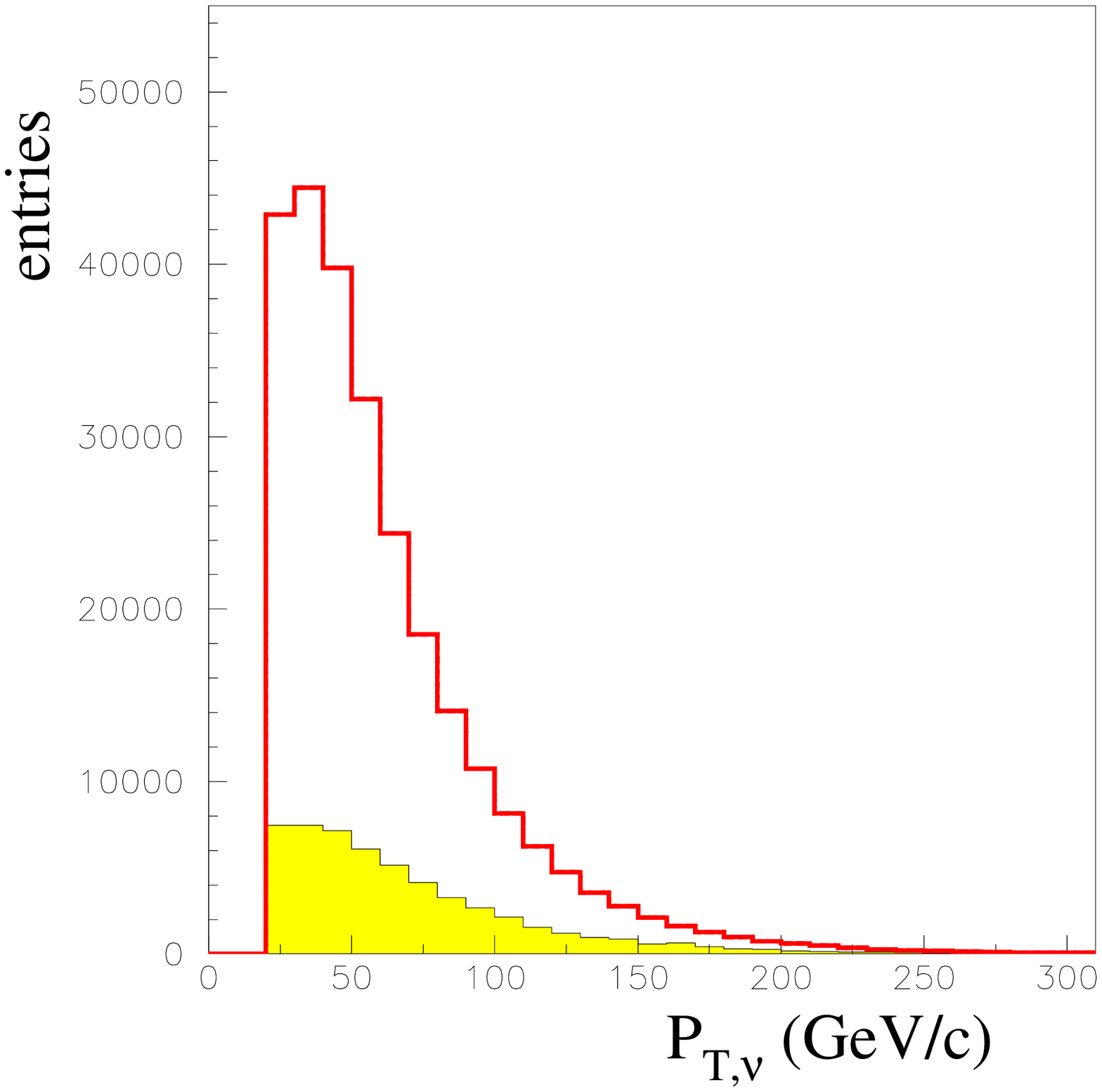,
height=5cm,clip=}} & \hspace{-6mm} &
\mbox{\epsfig{file=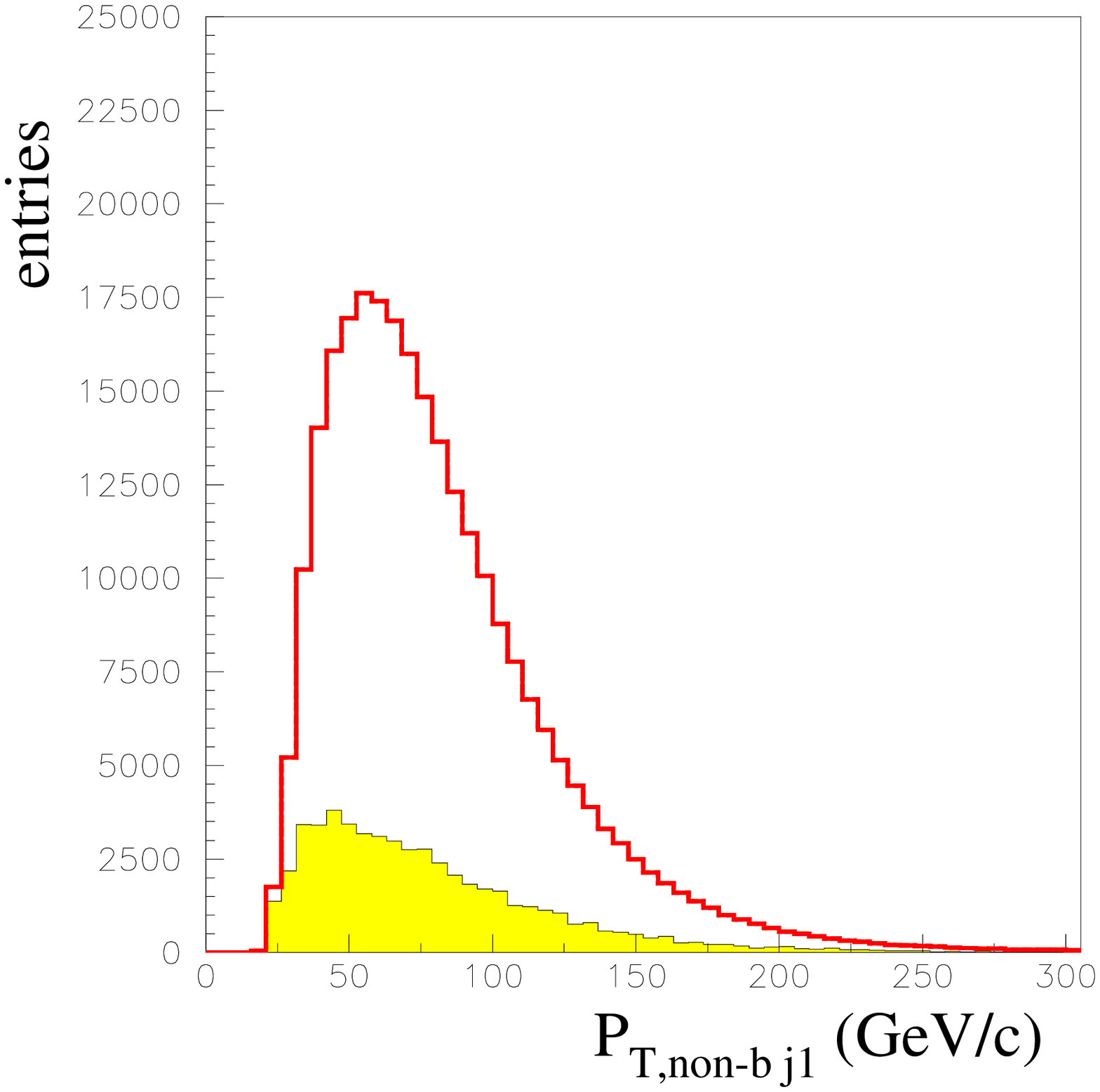,
height=5cm,clip=}} \\
(a) & \hspace{-6mm} & (b) & \hspace{-6mm} & (c) \\
\mbox{\epsfig{file=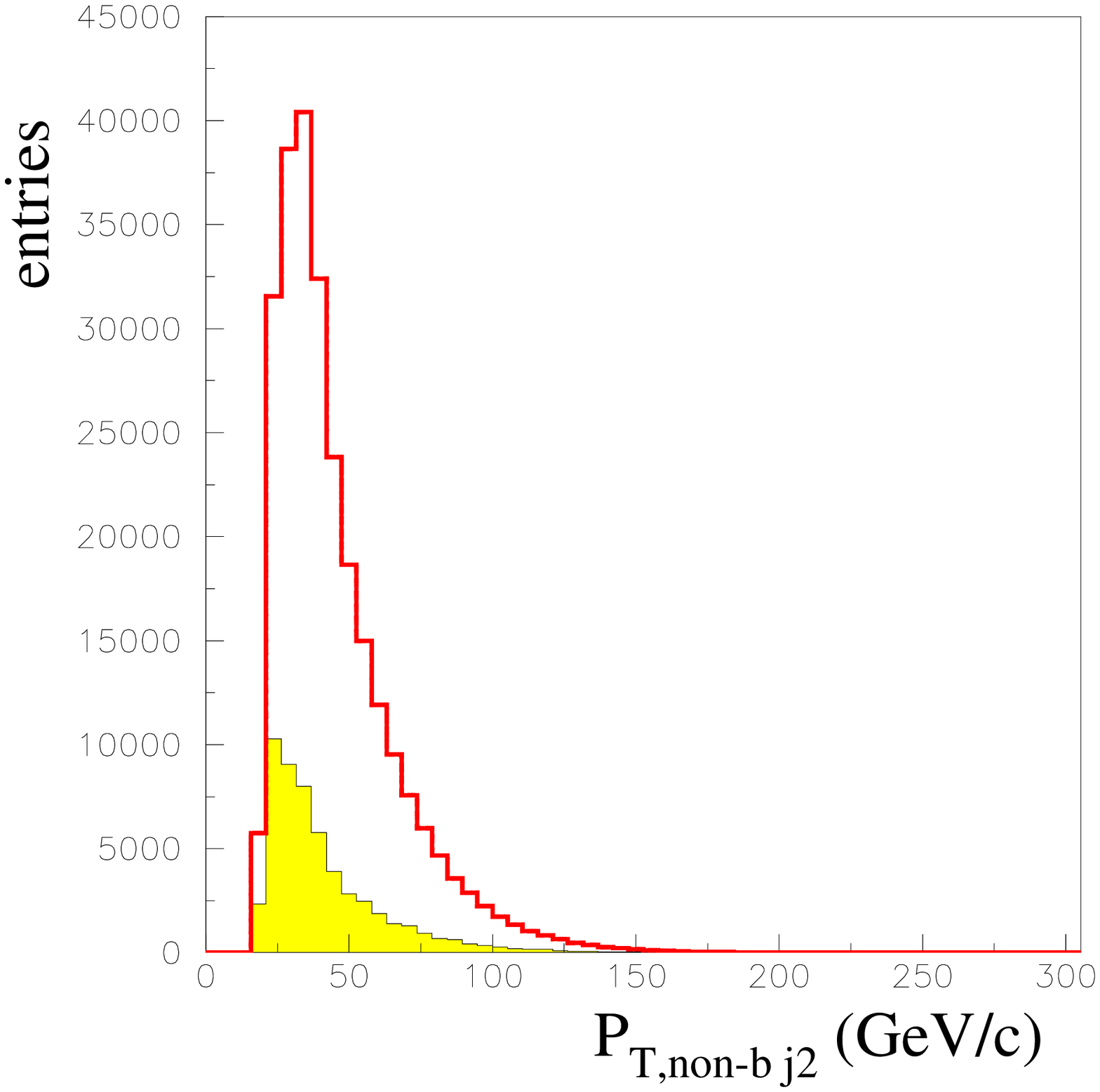,
height=5cm,clip=}} & \hspace{-6mm} &
\mbox{\epsfig{file=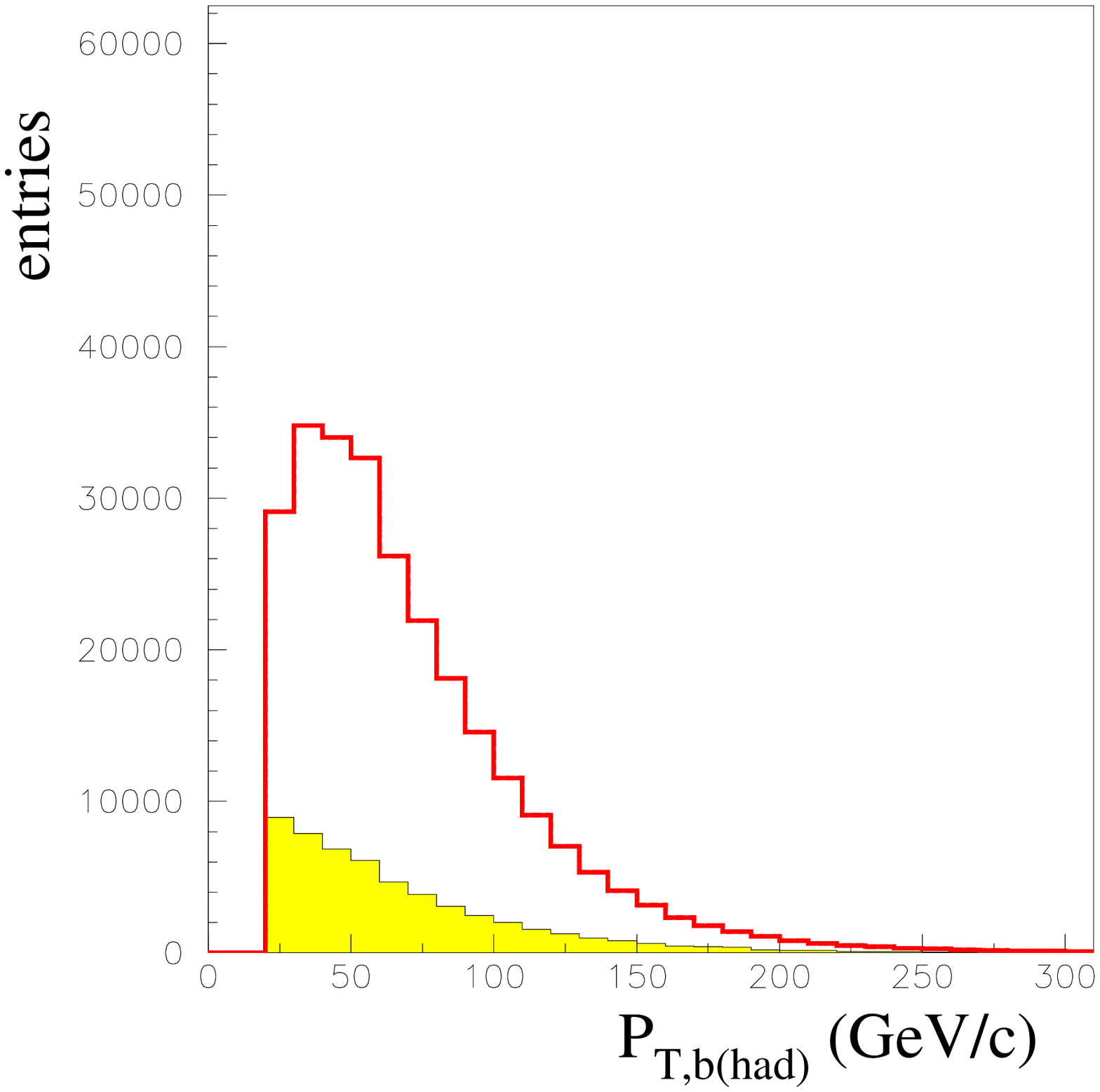,
height=5cm,clip=}} & \hspace{-6mm} &
\mbox{\epsfig{file=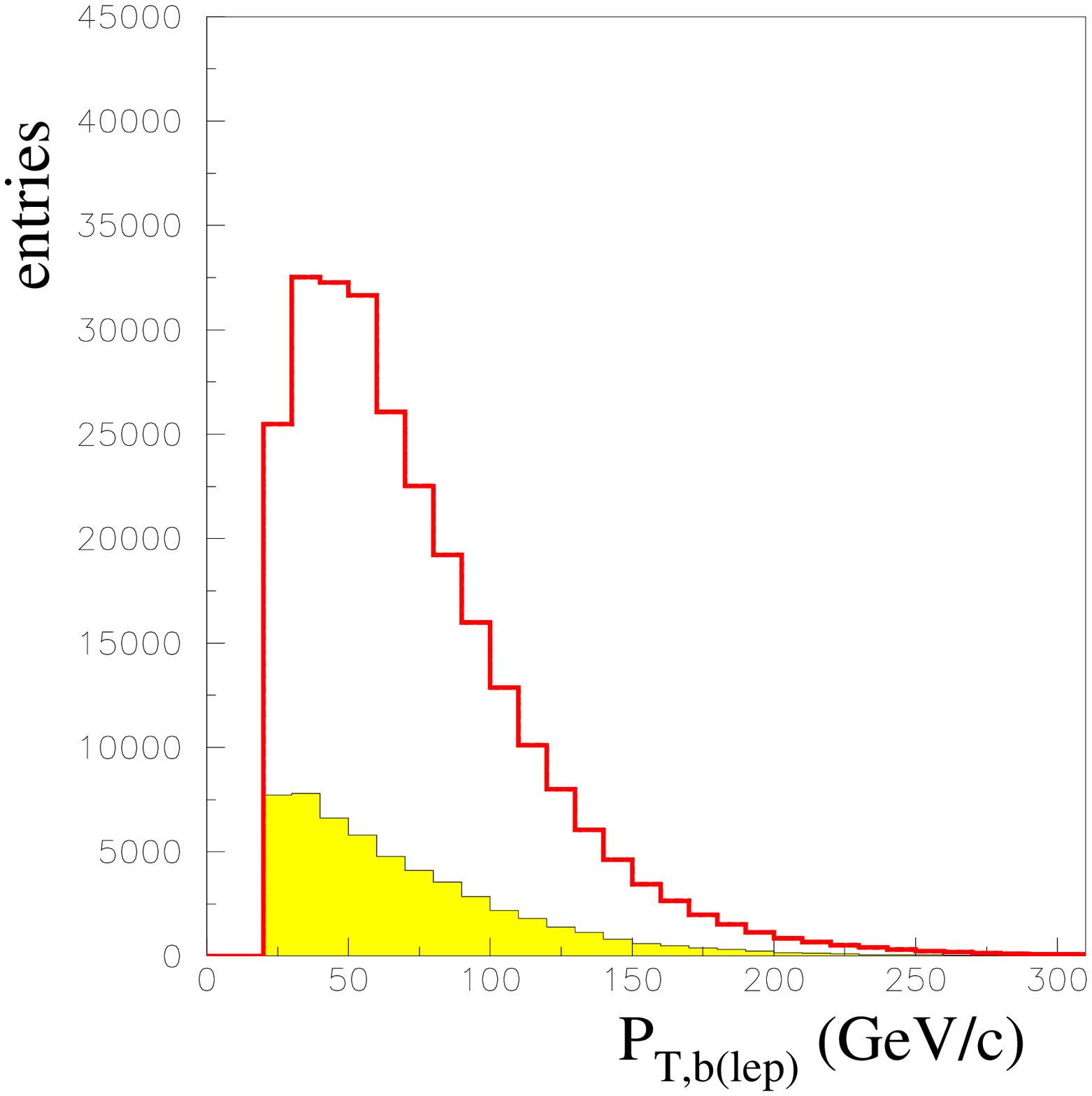,
height=5cm,clip=}} \\
(d) & \hspace{-6mm} & (e) & \hspace{-6mm} & (f) \\
\mbox{\epsfig{file=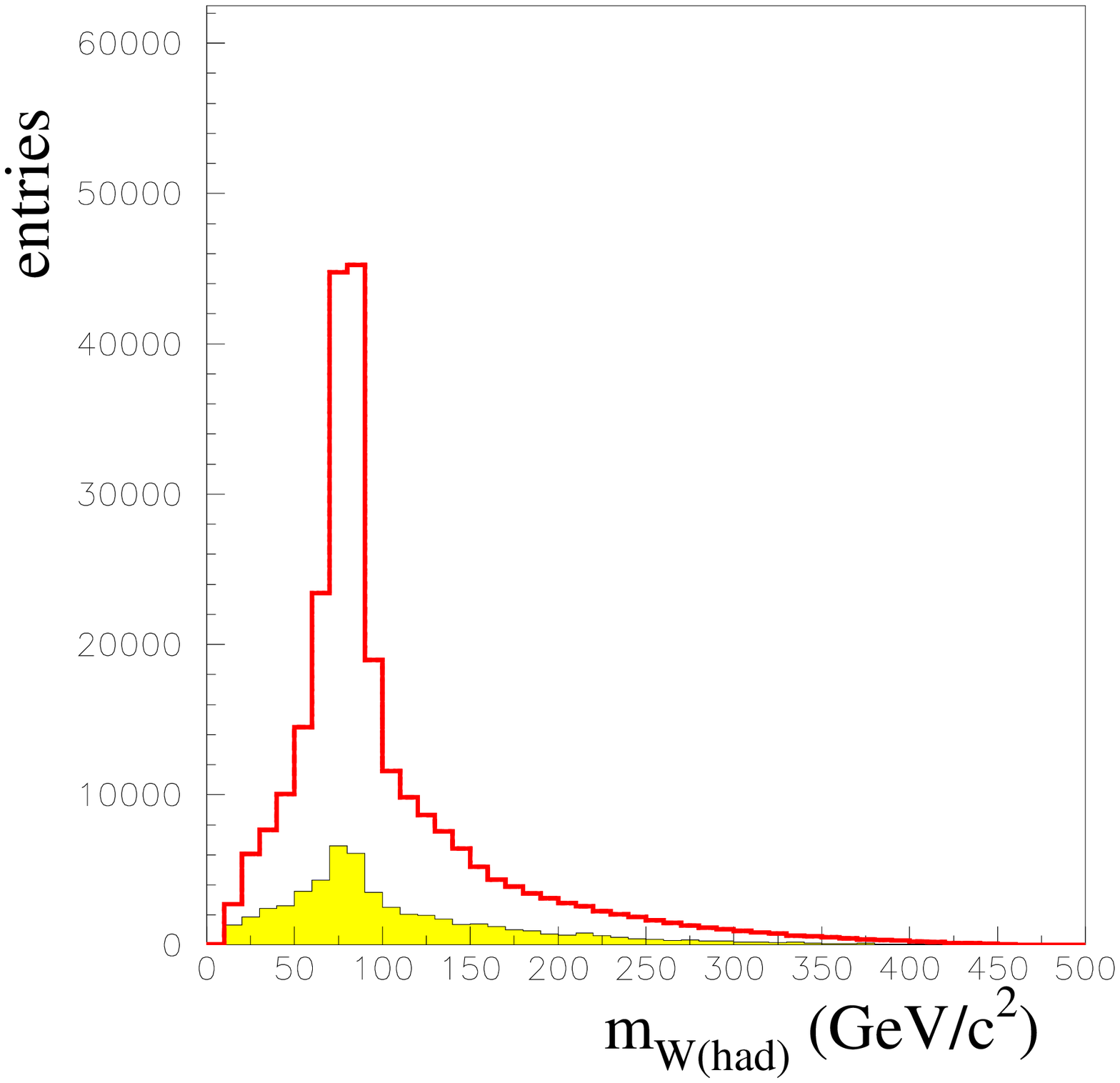,
height=5cm,clip=}} & \hspace{-6mm} &
\mbox{\epsfig{file=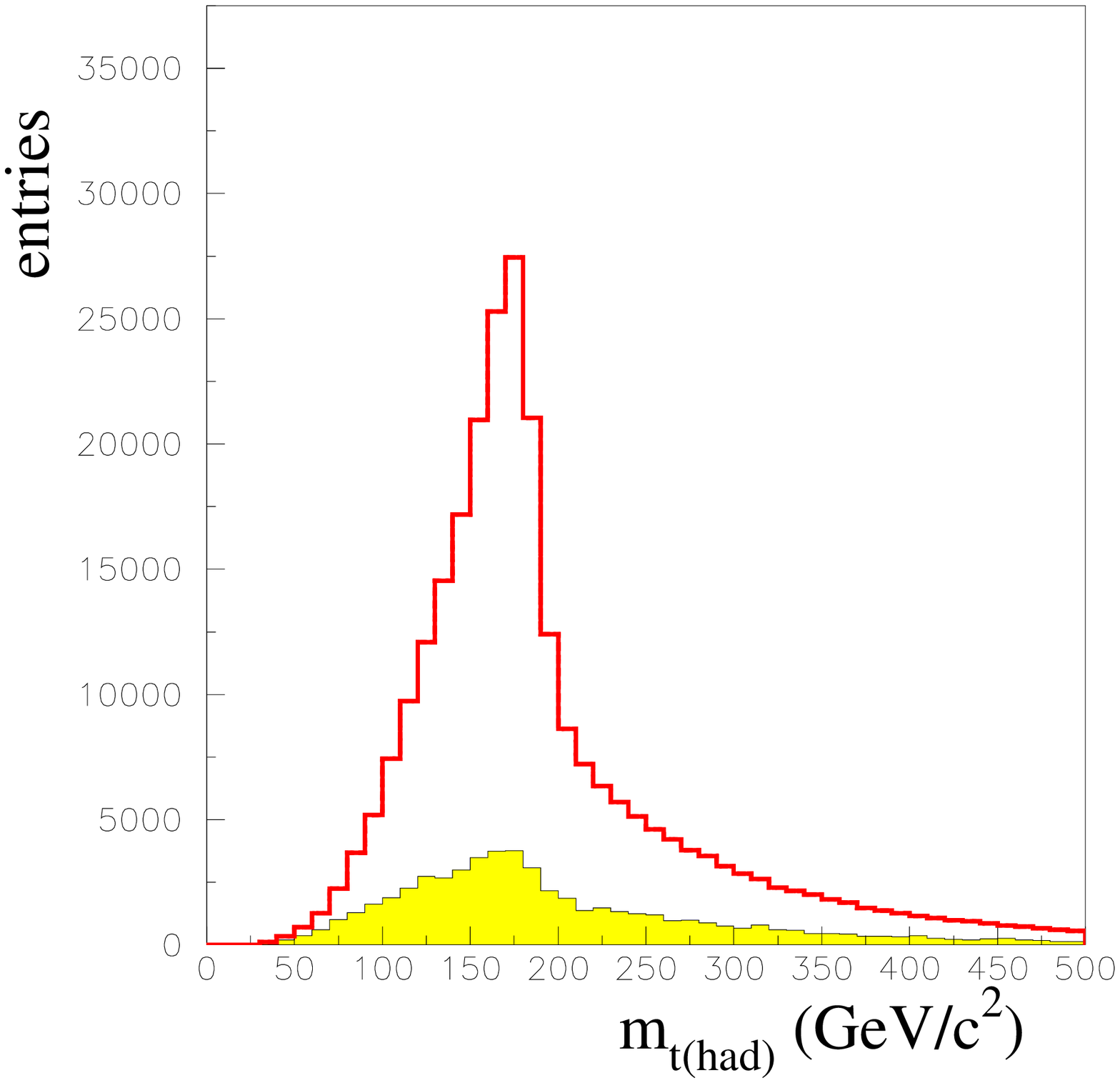,
height=5cm,clip=}} & \hspace{-6mm} &
\mbox{\epsfig{file=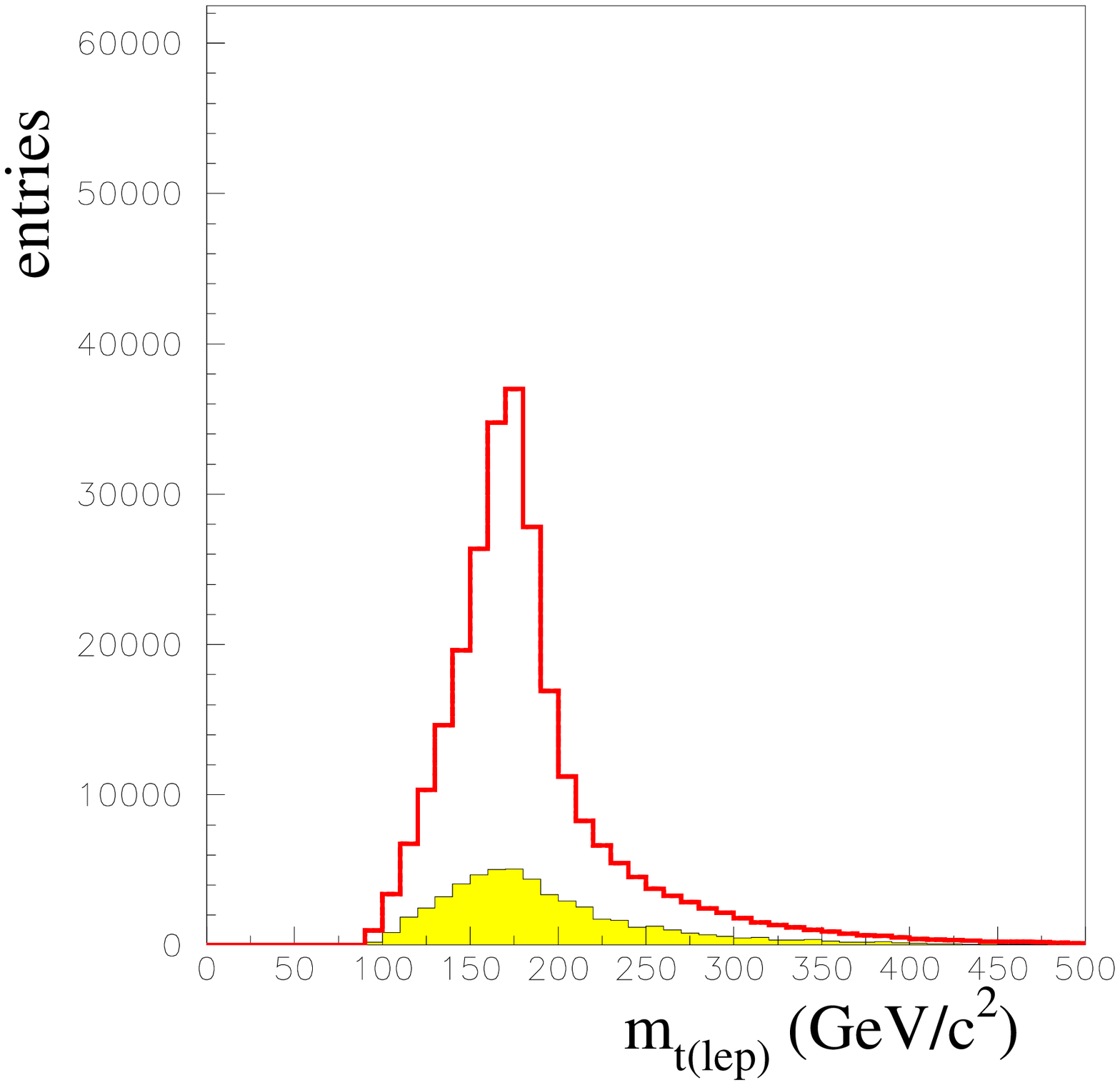,
height=5cm,clip=}} \\
(g) & \hspace{-6mm} & (h) & \hspace{-6mm} & (i) \\
\end{tabular}
\caption{Kinematical distributions at the pre-selection level for the
transverse momentum of the charged lepton (a), the neutrino (b), the 
$p_T$ of the two non $b$ jets used in the hadronic $W$ reconstruction (c),(d), the $b$ jet from the hadronic (e) and leptonic
(f) top quarks. Invariant mass distributions of the hadronic $W$ boson (g),
the hadronic top (h) and the leptonic top (i). The $t\bar t$ signal (full line)
and the SM backgrounds (shaded region) are normalised to $L=10$~fb$^{-1}$.}
\label{fig:kin}
\end{center}
\end{figure}

Signal events have a final state topology characterised by one isolated
lepton (the isolation criterium requires the absence of additional tracks with
$p_T > 10$~GeV inside a cone of $\Delta R = 0.4$ around the lepton direction),
at least four jets (among which exactly two must be tagged as $b$ jets) and
large transverse missing energy.
We apply a two-level probabilistic analysis,
based on the construction of a discriminant variable which uses the
full information of some kinematical properties of the event. 
In the first level (called the pre-selection), a cleaner sample is obtained
accepting events with: (i) exactly one charged lepton
with $p_T > 25$~GeV, $|\eta| < 2.5$; (ii) at least 4 jets with $p_T > 20$~GeV,
$|\eta| < 2.5$, two of them tagged as $b$ jets and at least two not $b$-tagged;
(iii) missing transverse momentum above 20~GeV. The number of signal and
background events (normalised to $L = 10$ fb$^{-1}$) and the signal efficiency
after the pre-selection are shown in the first column of Table~\ref{tab:prob}.
Distributions of relevant variables are presented in Fig.~\ref{fig:kin}.

\begin{table}[htb]
\begin{center}
\begin{tabular}{ccc}
Process & Pre-Selection & Final Selection \\
\hline
$t \bar t \to \ell \nu b \bar b q \bar q'$  & 262111 (11\%) & 220024 (9\%) \\
\hline
$t \bar t$ (other) & 36745 & 27060 \\
Single $t$         & 12410 &  7600 \\
$Z+$jets           &   566 &   253 \\
$W+$jets           &  3627 &  1307 \\
$WW$, $ZZ$, $ZW$   &   109 &    51 \\
\hline
total SM bkg.      & 53457 & 36271 \\
\hline
\end{tabular}
\caption{Number of signal $t \bar t \to \ell \nu b \bar b q \bar q'$ and
background events, normalised to $L=10$~fb$^{-1}$, after the pre-selection and
final selection. The $b \bar b$ background is negligible after selection.}
\label{tab:prob}
\end{center}
\end{table}

The hadronic $W$ reconstruction is done from the two non-$b$
jets with highest transverse momentum.
The invariant mass of these two jets is represented in Fig.~\ref{fig:kin} at
the pre-selection. The mass of the hadronic top , also shown
in Fig.~\ref{fig:kin}, is reconstructed as the
invariant mass of the hadronic $W$ and the $b$ jet (among the two with highest
$p_T$) closer to the $W$. The
leptonic $W$ momentum cannot be directly reconstructed due to the
presence of an undetected neutrino in the final state. Nevertheless,
the neutrino four-momentum can be estimated by assuming the transverse missing
energy to be the transverse neutrino momentum.
Its longitudinal component can
then be determined, with a quadratic ambiguity, by constraining the leptonic
$W$ mass (calculated as the invariant mass of the neutrino and the charged
lepton) to its known on-shell value $M_W \simeq 80.4$~GeV.
In order to solve the twofold quadratic ambiguity in the
longitudinal component it is required that the hadronic and the leptonic top
quarks have the minimum mass difference. The reconstructed mass of the
leptonic top is shown in Fig.~\ref{fig:kin} at the pre-selection.

\begin{figure}[htb]
\begin{center}
\mbox{\epsfig{file=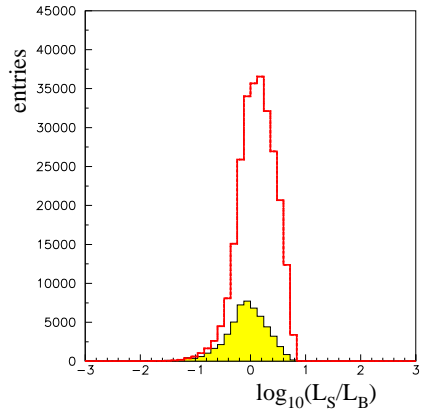,
height=6cm,clip=}}
\caption{Discriminant variable for the SM background (shaded region) and the
$t\bar t$ signal (full line), normalised to $L=10$~fb$^{-1}$.}
\label{fig:like}
\end{center}
\end{figure}

In the second
level (the final selection), for each event we construct signal and
background-like probabilities, 
$\mathcal P^\mathrm{signal}_i$ and $\mathcal P^\mathrm{back.}_i$, respectively,
using probability density functions (p.d.f.) built from relevant physical
variables:
\begin{itemize}
\item The hadronic $W$ mass.
\item The hadronic and leptonic top masses.
\item The transverse momentum of the $b$-jets associated to the hadronic
and the leptonic top quarks.
\item The transverse momentum of the jets used in the hadronic $W$
reconstruction.
\end{itemize}
These seven variables are shown in Fig.~\ref{fig:kin} (c-i).
Signal ($\mathcal{L}_S=\Pi^{n}_{i=1} \mathcal{P}^{signal}_i$) and
background ($\mathcal{L}_B=\Pi^{n}_{i=1} \mathcal{P}^{back.}_i$) likelihoods
(with $n=7$, the number of p.d.f.) are used to define a discriminant variable
$L_R= \log_{10} \mathcal{L}_S/\mathcal{L}_B$. This variable is shown in
Fig.~\ref{fig:like} for the signal and background. The final event selection
is done by applying a cut $L_R > -0.2$ on the discriminant variable, which
corresponds to the highest $S/\sqrt{B}$ ratio.
The number of background events (normalised to $L=10$ fb$^{-1}$) and signal
efficiency after the final selection are shown in the second column of
Table~\ref{tab:prob}.

Two more cut-based analyses, omitted here for brevity, have also been
performed. The results obtained depend more on the top mass reconstruction
method than on the type (cut-based or probabilistic) of analysis performed.
A detailed comparison of the three of them can be found in
Ref.~\cite{nota}, where it is shown that the probabilistic analysis presented
here gives the best results, with smaller systematic uncertainties.

\section{Experimental measurement of angular distributions and asymmetries}
\label{sec:4}

\begin{figure}[htb]
\begin{center}
\begin{tabular}{ccc}
\mbox{\epsfig{file=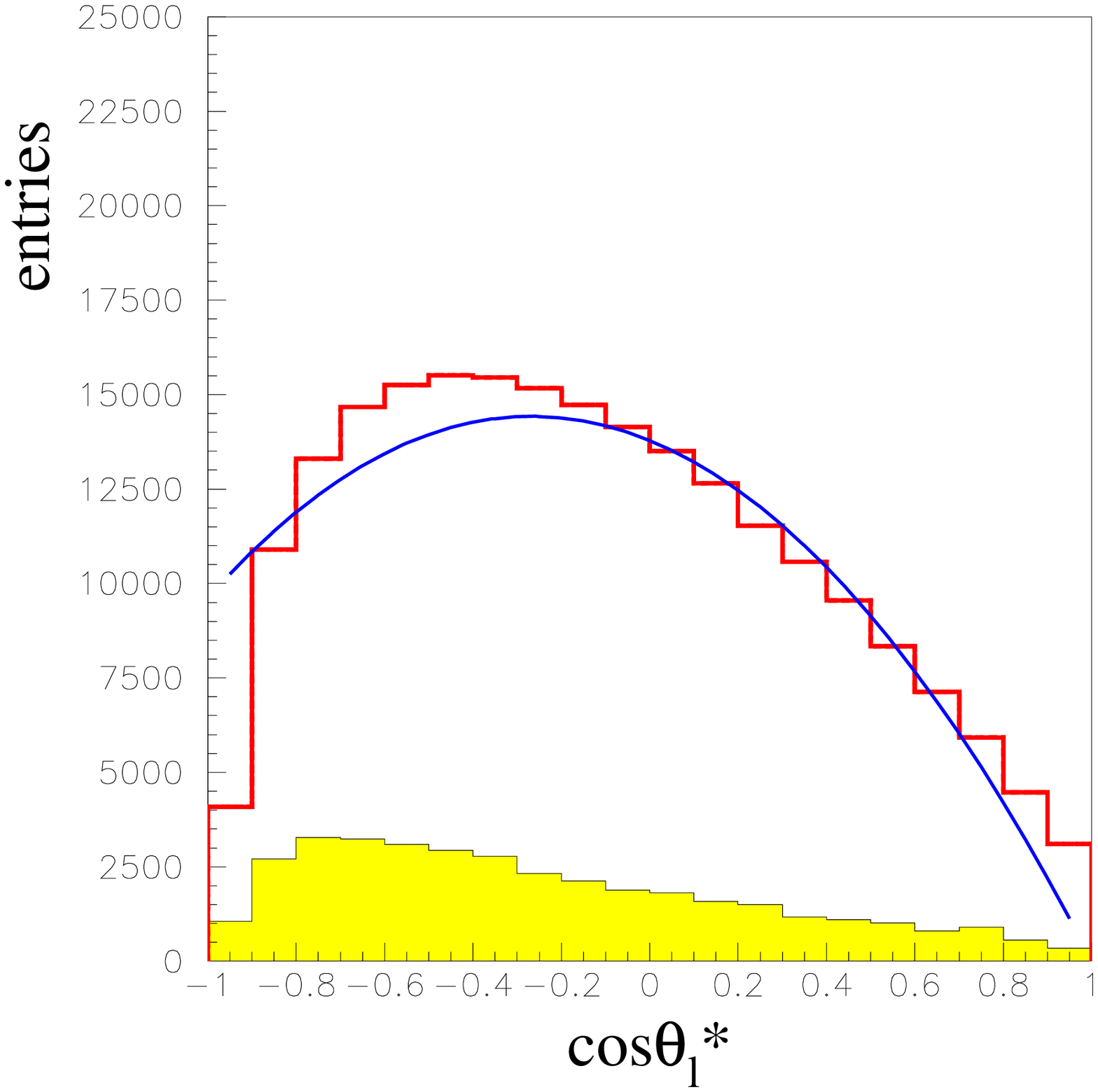,
height=6cm,clip=}} & \hspace{-6mm} &
\mbox{\epsfig{file=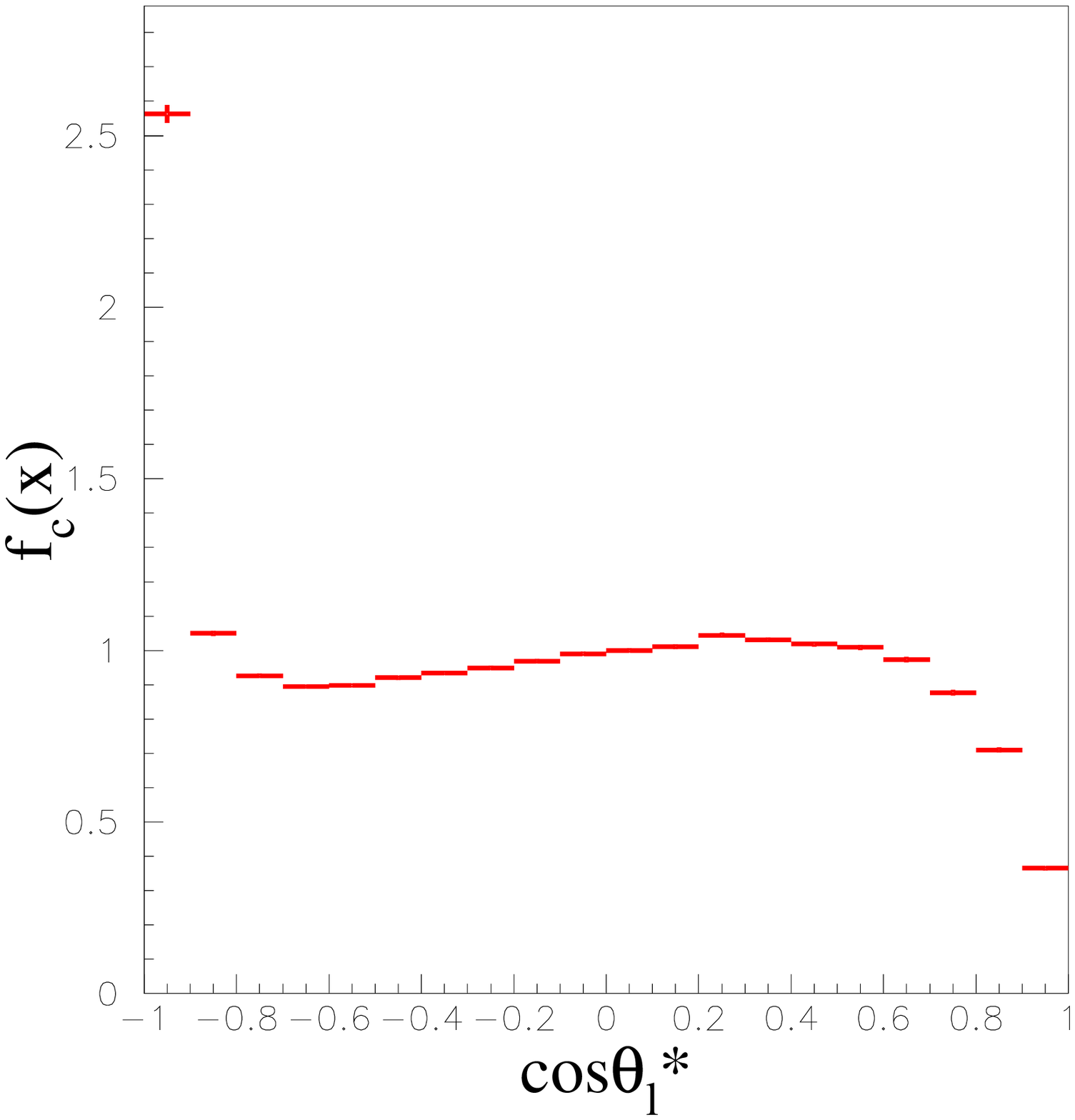,
height=6cm,clip=}} \\
(a) & \hspace{-6mm} & (b) 
\end{tabular}
\caption{Simulated $\cos \thlw$ distribution (a) and its correction function
(b). In the first plot the $t\bar t$ signal (full line)
and the SM backgrounds (shaded region) are normalised to $L=10$~fb$^{-1}$.}
\label{fig:dist}
\end{center}
\end{figure}

The experimentally observed $\cos \thlw$ distribution, which includes the
$t \bar t$ signal as well as the SM backgrounds, is affected by
detector resolution, $t \bar t$ reconstruction and selection criteria.
In order to recover the theoretical distribution, it is necessary to: (i)
subtract the background; (ii) correct for the effects of the detector,
reconstruction, etc. For this purpose, we use two different sets of signal and
background event samples:
one ``experimental'' set, which simulates a possible experimental result, and
one ``reference'' set, which is used to
parameterise the effects mentioned and correct the previous sample.
The procedure is as follows. After subtracting reference background
samples, the ``experimental'' distribution 
 is multiplied by a correction
function $f_c$ in order to recover the theoretical one expected in the
SM.\footnote{Correction functions are determined assuming that the charged
lepton distribution corresponds to the SM one. In case that a deviation from SM
predictions (corresponding to anomalous couplings) is found, the correction
function must be modified accordingly, and the theoretical distribution
recalculated in an iterative process. These
issues have been analysed in detail in Ref.~\cite{marsella}, where it is shown
that this process quickly converges.}
The correction function is calculated, for each bin of the $\cos \thlw$
distribution, dividing the number of events at the generator level by the
number of events after the event selection, using the reference sample.
The ``experimental'' $\cos \thlw$ distribution obtained after the simulation
is shown in Fig.~\ref{fig:dist}, together with the correction function obtained
from the reference sample.
The asymmetries are measured with a simple counting of the number of events
below and above a specific value of $\cos \thlw$ as in 
Eq.~(\ref{ec:apm}). The procedure to correct for
detector and reconstruction effects is basically the same, but with the $\cos
\thlw$ distribution divided into two or three bins. This has the advantage that the
asymmetry measurements are not biased by the extreme values of the angular
distributions, where correction functions largely deviate from unity and special
care is required (see Fig.~\ref{fig:dist}).
The helicity fractions and ratios obtained from a fit to the corrected
distribution, as well as the angular asymmetries $\afb$, $A_\pm$, are
collected in Table~\ref{tab:stat}, with their statistical uncertainties. For
easy comparison, we also include the theoretical values obtained at the
generator level.

\begin{table}[htb]
\begin{center}
\begin{tabular}{cccc|cc|ccc}
      & $\fz$ & $\fm$ & $\fp$                & $\rhm$ & $\rhp$               & $\afb$  & $\Ap$  & $\Am$ \\
\hline
Th.   & 0.703 & 0.297 & $3.6 \times 10^{-4}$ & 0.423  & $5.1 \times 10^{-4}$ & -0.2220 & 0.5493 & -0.8402 \\
Rec.  & 0.700 & 0.299 & 0.0006               & 0.4274 & 0.0004               & -0.2231 & 0.5472 & -0.8387 \\
$\Delta$stat.
      & 0.003 & 0.003 & 0.0012               & 0.0080 & 0.0021               &  0.0035 & 0.0032 &  0.0018 \\
\hline
\end{tabular}
\caption{Theoretical and reconstructed values of helicity fractions, helicity
ratios and angular asymmetries, with their statistical errors for $L=10$
fb$^{-1}$.}
\label{tab:stat}
\end{center}
\end{table}

Due to the excellent statistics achievable at LHC, which is apparent in
Table~\ref{tab:stat}, systematic errors play a
crucial role in the measurement of angular distributions and asymmetries
for a luminosity of 10 fb$^{-1}$ or larger. A
thorough discussion of the different systematic uncertainties in the
determination of the correction functions is therefore compulsory. We estimate
the systematic errors in the observables studied (asymmetries, helicity
fractions and ratios) by
calculating them with various reference samples and observing the differences
obtained. In some cases the estimates are conservative, and they are taken as
a reference for better comparison with previous analyses \cite{marsella}.
We consider uncertainties originating from:

{\it Monte Carlo generator:} The correction functions obtained from a sample
generated with TopReX are applied to a sample generated with
ALPGEN~\cite{alpgen}. The
difference between the values obtained at the generator level and after the
simulation is considered as systematic uncertainty.

{\it Structure functions:} The correction functions obtained from a reference
sample generated with CTEQ5L PDFs are applied to samples
generated with CTEQ6L and MRST2001 PDFs in order to estimate the effects 
on the correction functions, and thus on
the observables. The most significant deviations found are considered as the
systematic error associated to the structure functions. 

{\it Top mass dependence:} Samples corresponding to top masses of 170, 175 and
180~GeV are simulated, and the influence of $m_t$ on the values obtained for
the corrected observables (using correction functions for $m_t = 175$~GeV) is
determined. The systematic error used here is obtained from a linear fit of the
values found corresponding to a top mass uncertainty of 2~GeV.

{\it ISR and FSR:} Their effect is studied following
Ref.~\cite{Borjanovic:2004ce}. An event sample is created
in which ISR and FSR are switched off in the
event simulation. We compare the results of the reference sample (with ISR and
FSR) with those obtained adding to it a normalised fraction of the sample
without ISR nor FSR (from 0\% to 25\%, in steps of 5\%).
The values obtained for the observables are fitted with a
linear function and the systematic error is
considered as the effect of the presence of 20\%
(a conservative estimate of our level of knowledge of ISR and FSR) of the
sample without ISR, FSR.

{\it $b$ jet tag efficiency:} The value of the $b$ jet tag efficiency (and the
corresponding $c$ jet and light jet rejection factors) is varied from 50 to
70\%, in steps of 5\%, and the values obtained for the observables are fitted
with a linear function
The systematic error is considered as the effect on the observables of a
variation of 5\% in the $b$ jet tagging efficiency, as compared with the
standard value of 60\%.

{\it $b$ jet energy scale:} The value of the $b$ jet energy scale is changed
from -5 to +5\%, and the values obtained for the observables are fitted with a
linear function. The systematic error is considered as the effect
of a variation of 3\% in the $b$ jet energy scale.

{\it Light jet energy scale:} The value of the energy scale of the light jets
is changed from -3 to +3\%, and the values obtained for the observables are
fitted with a linear function.
The systematic error is considered as the effect of
a variation of 1\% in the energy scale of the light jets.

{\it Background:} The background (as obtained from the reference sample)
subtracted to the selected sample is varied from -25 to 25\%, in steps of 5\%,
and the values obtained for the observables are fitted with a linear function.
The systematic error is considered as the effect of a
variation of 10\% on the background level (which takes into account the
uncertainties in the cross-sections).

{\it Pile-up:} The effect of pile-up events (2.3 events in average) is
studied by comparing the values of the observables obtained with and without
adding pile-up events.

{\it $b$ quark fragmentation:} The parameter $\epsilon_b$ in the Peterson
parameterisation for $b$ quark fragmentation is changed from 
-0.006 to -0.0035, and the values obtained for the observables compared. The
difference is considered as systematic error \cite{Borjanovic:2004ce}. 

The systematic errors in each observable, resulting from these theoretical and
simulation uncertainties, are collected in Table~\ref{tab:sys}. It can be
observed that $\rhp$ and $\Am$ have very small total systematic errors.
In the case of $\rhp$, the improvement over $\fp$ is due to the cancellation
of some of the systematic errors in the ratio, while the opposite happens
in the case of $\rhm$, compared to $\fm$.

\begin{table}[htb]
\begin{center}
\begin{tabular}{cccc|cc|ccc}
Source & $\fz$ & $\fm$ & $\fp$ & $\rhm$ & $\rhp$ & $\afb$ & $\Ap$ & $\Am$ \\
\hline
MC generator & 0.0002 & 0.0002 & 0.0004 & 0.0006 & 0.0000 & 0.0035 & 0.0015 & 0.0006 \\
PDFs         & 0.0032 & 0.0022 & 0.0009 & 0.0046 & 0.0008 & 0.0021 & 0.0005 & 0.0014 \\
Top mass     & 0.0065 & 0.0060 & 0.0006 & 0.0124 & 0.0007 & 0.0034 & 0.0039 & 0.0005 \\
ISR+FSR      & 0.0116 & 0.0113 & 0.0003 & 0.0218 & 0.0001 & 0.0046 & 0.0049 & 0.0011 \\
$b$ tag eff. & 0.0065 & 0.0062 & 0.0003 & 0.0126 & 0.0003 & 0.0039 & 0.0046 & 0.0004 \\
$E_b$ scale  & 0.0028 & 0.0030 & 0.0002 & 0.0061 & 0.0002 & 0.0021 & 0.0017 & 0.0005 \\
$E_j$ scale  & 0.0034 & 0.0037 & 0.0002 & 0.0074 & 0.0002 & 0.0038 & 0.0023 & 0.0014 \\
Back.        & 0.0001 & 0.0000 & 0.0000 & 0.0001 & 0.0000 & 0.0001 & 0.0000 & 0.0001 \\
Pile-up      & 0.0091 & 0.0086 & 0.0005 & 0.0175 & 0.0002 & 0.0080 & 0.0051 & 0.0006 \\
$b$ frag.    & 0.0049 & 0.0037 & 0.0012 & 0.0078 & 0.0011 & 0.0045 & 0.0000 & 0.0012 \\
\hline
Total $\Delta$sys.
             & 0.0189 & 0.0178 & 0.0018 & 0.0356 & 0.0016 & 0.0130 & 0.0099 & 0.0028 \\
\hline
\end{tabular}
\caption{Sources of systematic errors in the determination of helicity
fractions, helicity ratios and angular asymmetries.}
\label{tab:sys}
\end{center}
\end{table}

The reduction of systematic errors compared to previous analyses deserves
an explanation. In this
analysis the $W$ helicity fractions and ratios 
are obtained by fitting the angular distribution from $-0.99$ to $+0.99$,
and it should be stressed 
that a dependence of the systematic error with the range of the fit has been observed. If the fit is performed between
$-0.89$ and $+0.89$, the systematic errors on $\fz$, $\fm$ and $\fp$ are respectively $0.0206$, $0.0188$ and $0.0033$
(in good agreement with the results of Ref.~\cite{marsella}).
However, if the fit is performed in the range [$-0.89$,$+0.99$] the results are respectively $0.0190$, $0.0182$ and 
$0.0017$, still in good agreement with the values on Table~\ref{tab:sys}.
This implies that the correct reconstruction of the most extreme bins of the angular distribution is of utmost importance 
in order to control the error associated to the $W$ polarisation measurements, if the fitting method is used.
In the case of the asymmetries, for $\Apm$ the smaller
errors are due to the greater stability of these measurements, obtained by
counting events, compared to observables obtained from a fit to the
$\cos \thlw$ distribution. We point out that the selection of $z$ for the
definition of $\Apm$ in Eqs.~(\ref{ec:apm}) has not been optimised
in order to achieve smaller systematic errors. Instead, these asymmetries
have been defined in a simple way which allows to reconstruct easily the
helicity fractions, using Eqs.~(\ref{ec:inv}).
The results of our simulation, including statistical and systematic
uncertainties, are summarised in Table~\ref{tab:results}.

\begin{table}[htb]
\begin{center}
\begin{tabular}{crll}
Observable & \multicolumn{3}{c}{Result} \\
\hline
$\fz$   & $0.700$   & $ \pm 0.003\,\mathrm{(stat)}$  & $ \pm 0.019\,\mathrm{(sys)}$ \\
$\fm$   & $0.299$   & $ \pm 0.003\,\mathrm{(stat)}$  & $ \pm 0.018\,\mathrm{(sys)}$ \\
$\fp$   & $0.0006$  & $ \pm 0.0012\,\mathrm{(stat)}$ & $ \pm 0.0018\,\mathrm{(sys)}$ \\
$\rhm$  & $0.4274$  & $ \pm 0.0080\,\mathrm{(stat)}$ & $ \pm 0.0356\,\mathrm{(sys)}$ \\
$\rhp$  & $0.0004$  & $ \pm 0.0021\,\mathrm{(stat)}$ & $ \pm 0.0016\,\mathrm{(sys)}$ \\
$\afb$  & $-0.2231$ & $ \pm 0.0035\,\mathrm{(stat)}$ & $ \pm 0.0130\,\mathrm{(sys)}$ \\
$\Ap$   & $0.5472$  & $ \pm 0.0032\,\mathrm{(stat)}$ & $ \pm 0.0099\,\mathrm{(sys)}$ \\
$\Am$   & $-0.8387$ & $ \pm 0.0018\,\mathrm{(stat)}$ & $ \pm 0.0028\,\mathrm{(sys)}$ \\
\hline
\end{tabular}
\caption{Summary of the results obtained from the simulation for the
observables studied, including statistical and systematic uncertainties.}
\label{tab:results}
\end{center}
\end{table}

\section{Limits on anomalous couplings}
\label{sec:5}

With the results obtained in the previous section, summarised in
Table~\ref{tab:results}, and the parametric dependence of the observables on
$\vr$, $\gl$ and $\gr$ implemented in the computer program {\tt TopFit}
\cite{paper1}, constraints on
the latter can be set. Naively, to obtain the $1\sigma$ limit on a
coupling $x=\vr,\;\gl,\;\gr$ derived from the measurement of some observable
$O$,
one would simply find the values of $x$ for which $O$ deviates $1\sigma$ from
its central value.\footnote{This is the procedure originally followed in
our previous work \cite{nota}, as well as in Ref.~\cite{marsella}:
For an observable $O$ and a coupling $x$, intersecting the plot of $O(x)$
with the two horizontal lines $O = O_\text{exp} \pm \Delta O$, which correspond
to the $1 \sigma$ variation of $O$, gives the pretended $1 \sigma$ interval
on $x$.}
Nevertheless, due to the quadratic dependence of the observables on
$\vr$ and $\gl$ near the SM point $\vr = \gl = 0$,
this procedure leads to overcoverage
of the obtained confidence intervals \cite{paper1}, because
their p.d.f. is not Gaussian even if the p.d.f. of the observable $O$ is.
In order to obtain the limits on an anomalous coupling $x$, given by the
measurement of an observable $O$,  we determine the p.d.f. of $x$ numerically,
using the acceptance-rejection method: we iteratively
(i) generate a random value (with uniform probability) $x_i$ within a suitable
interval; (ii) evaluate the probability of $O(x_i)$, given by the p.d.f. of $O$;
(iii) generate an independent
random number $r_i$ (with uniform probability); and (iv) accept the value $x_i$
if the probability of $O(x_i)$ is larger than $r_i$. The resulting set of values
$\{x_i\}$ is distributed according to the p.d.f. of $x$ given by the measurement
of $O$. The determination of a
central interval with a given CL $\gamma$ is done
numerically, requiring: (a) that it contains a fraction $\gamma$ of the total
number of values $\{x_i\}$; (b) that is central, {\em i.e.} fractions
$(1-\gamma)/2$ of the values generated are on each side of the interval.

For $x=\gr$ this method gives results very similar to the intersection method in
Refs.~\cite{marsella,nota},
whereas for $\vr$ and $\gl$ the confidence intervals found are 20\% and 30\%
smaller, respectively.
The $1\sigma$ limits derived from the measurement of each observable
are collected in Table~\ref{tab:lim1}, assuming only one
nonzero coupling at a time.  We notice the improvement in
sensitivity brought by the new observables $\rhpm$ and $\Apm$:
the best limits on $\vr$ and $\gl$ are obtained
from the measurement of $\rhp$, improving the limits from $\fp$ by a factor of
1.13, and the best limits on $\gr$ are provided by $\Ap$, improving the limits
from $\fm$ by a factor of 1.34. This is due to the smaller (systematic plus statistical)
uncertainties of these new observables and their stronger dependence on
anomalous couplings.

\begin{table}[htb]
\begin{center}
\begin{tabular}{cccc}
          &       $\vr$        &       $\gl$         &     $\gr$          \\[-1mm]
          &   ($\gl=\gr=0$)    &   ($\vr=\gr=0$)     &   ($\vr=\gl=0$)    \\
\hline
$\fz$     & --                 & $[-0.133,0.102]$    & $[-0.0315,0.0219]$ \\
$\fm$     & $[-0.196,0.186]$   & $[-0.167,0.136]$    & $[-0.0293,0.0212]$ \\
$\fp$     & $[-0.0373,0.1070]$   & $[-0.0491,0.0169]$  & --                 \\
$\rhm$    & $[-0.254,0.206]$   & --                  & $[-0.0275,0.0227]$ \\
$\rhp$    & $[-0.0282,0.0987]$ & $[-0.0455,0.0129]$  & --                 \\
$\afb$    & $[-0.118,0.148]$   & $[-0.0902,0.0585]$  & $[-0.0268,0.0227]$ \\
$\Ap$     & $[-0.140,0.146]$   & $[-0.112,0.0819]$   & $[-0.0213,0.0164]$ \\
$\Am$     & $[-0.0664,0.120]$  & $[-0.0620,0.0299]$  & $[-0.0166,0.0282]$ \\
\hline
\end{tabular}
\caption{Limits on anomalous couplings obtained by the measurement of the
observables in the left column, with the constraint that only one non-standard
coupling is allowed to be nonzero at a
time. Dashes are shown where there is no significant sensitivity.}
\label{tab:lim1}
\end{center}
\end{table}

These limits can be further improved by combining the measurements of the four
observables $\rhpm$ and $\Apm$, including their correlations.
We point out that the correlations among $\Apm$, $\rhpm$
do depend (as they must) on the method followed to extract these observables
from experimental data. In our analysis $\Apm$ are obtained by a simple event
counting above and below a specific value of $z = \cos \thlw$, while $\rhpm$
are obtained from a fit to the $\cos \thlw$ distribution, divided in 20
bins. The correlations among these observables are derived as follows.
We use a set of hypothetical ``experimental measurements'', in which each
element of the set is a binned $\cos \thlw$ distribution, as it
would be experimentally obtained after correcting for detector effects. For
each ``measurement'',
the number of events in each $\cos \thlw$ bin is obtained randomly using a
Gaussian distribution centered at the expected SM value.
We then calculate the average on this set, denoted by $\langle \cdot \rangle$,
of the ten independent products of observables
$\langle \Ap^2 \rangle$, $\langle \Ap \Am \rangle$, 
$\langle \Ap \rhm \rangle$, etc.,
where $\Apm$, $\rhpm$ are extracted from the
$\cos \thlw$ distribution as indicated above.\footnote{Since the four
observables $\Apm$, $\rhpm$ are obtained from the corrected $\cos \thlw$
distribution, there is no need to know the full kinematics of the $t \bar t$
event in order to determine their statistical correlation.
On the other hand, if we are interested in, for example, the correlation
between one of these observables and a top-antitop spin asymmetry,
the full $t \bar t$
kinematics is needed. In the latter case, systematic errors can possibly
influence the determination of the correlations.}
The resulting correlation matrix is shown in Table~\ref{tab:corr}.
The correlations among $\Apm$ and $\rhpm$ obtained are not affected
by systematic uncertainties, as long as these do not significantly distort
the shape of the $\cos \thlw$ distribution with respect to the SM one.

\begin{table}[htb]
\begin{center}
\begin{tabular}{ccccc}
& $\Ap$ & $\Am$ & $\rhm$ & $\rhp$ \\
$\Ap$  & 1       & 0.1587   & -0.8222  & -0.1232 \\
$\Am$  & 0.1587  & 1        & -0.08583 & 0.5688 \\
$\rhm$ & -0.8222 & -0.08583 & 1        & 0.3957 \\
$\rhp$ & -0.1232 & 0.5688   & 0.3957   & 1 \\
\hline
\end{tabular}
\caption{Correlation matrix for $\Apm$, $\rhpm$.}
\label{tab:corr}
\end{center}
\end{table}

When the four observables $\Apm$ and $\rhpm$ are combined the assumption
that only one coupling is nonzero can be relaxed.
However, if $\vr$ and $\gl$ are simultaneously allowed to be arbitrary,
the limits on them are very loose and correlated, because for fine-tuned values of
these couplings their effects on helicity fractions cancel to a large extent.
In this way, values $O(0.4)$ of $\vr$ and $\gl$ are possible yielding minimal
deviations on the observables studied.
Therefore, in our combined limits, which are presented in Table~\ref{tab:lim2},
we require that either $\vr$ or $\gl$ vanishes. Limits for both $\vr$, $\gl$
nonzero require additional observables beyond the ones directly related
to $W$ helicity fractions, and will be presented elsewhere.

\begin{table}[htb]
\begin{center}
\begin{tabular}{cccc}
& $\vr$ & $\gl$ & $\gr$ \\
\hline
$\Apm$, $\rhpm$
        & $[-0.0195,0.0906]$ & $\times$            & $\times$ \\
$\Apm$, $\rhpm$
        & $\times$           & $[-0.0409,0.00926]$ & $\times$ \\
$\Apm$, $\rhpm$
        & $\times$           & $\times$            & $[-0.0112,0.0174]$ \\
$\Apm$, $\rhpm$
        & $\times$           & $[-0.0412,0.00944]$ & $[-0.0108,0.0175]$ \\
$\Apm$, $\rhpm$
        & $[-0.0199,0.0903]$ & $\times$            & $[-0.0126,0.0164]$ \\
\hline
\end{tabular}
\caption{Limits on anomalous couplings obtained from the combined measurement
of $\Apm$, $\rhpm$.
In each case, the couplings which are fixed to be zero are denoted by a cross.}
\label{tab:lim2}
\end{center}
\end{table}

\begin{figure}[htb]
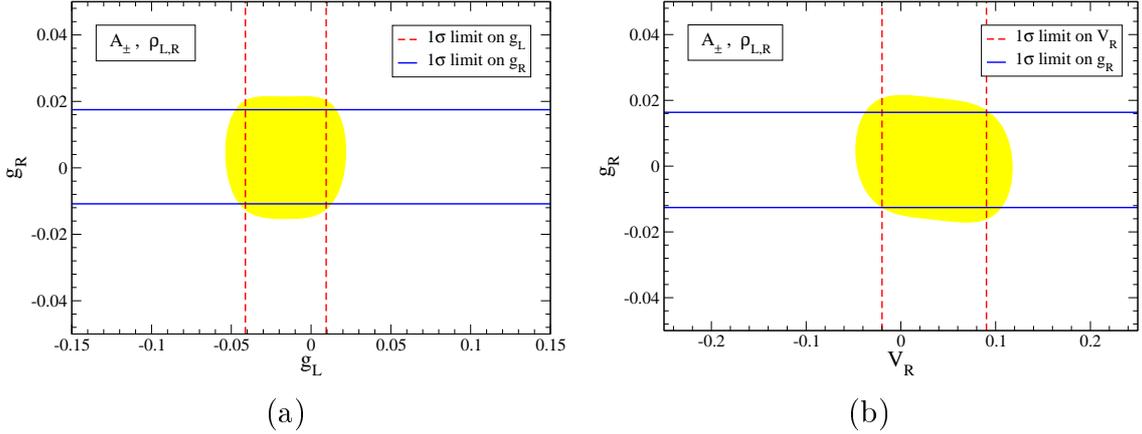

\begin{center}
\begin{tabular}{cc}
\mbox{\epsfig{file=Figs/reg-glgr-Arho.eps,height=5cm,clip=}} &
\mbox{\epsfig{file=Figs/reg-vrgr-Arho.eps,height=5cm,clip=}} \\
(a) & (b)
\end{tabular}
\caption{68.3\% CL confidence regions on anomalous couplings: $\gl$ and $\gr$,
for $\vr=0$ (a); $\vr$ and $\gr$, for $\gl = 0$ (b).
The $1\sigma$ combined limits in Table~\ref{tab:lim2} are also displayed.}
\label{fig:reg}
\end{center}
\end{figure}

For completeness, and to compare with previous literature we also present the
$2\sigma$ limits on non-standard couplings when only one of them is nonzero,
\begin{align}
\vr\;(2\sigma) & \quad [-0.0566,0.128] & (\gl=\gr=0) \,, \notag \\
\gl\;(2\sigma) & \quad [-0.0579,0.0258] & (\vr=\gr=0) \,, \notag \\
\gr\;(2\sigma) & \quad [-0.0260,0.0312] & (\vr=\gl=0) \,.
\end{align}
A significant improvement, by factors of 3.25, 3.1 and 1.4, respectively,
is obtained with the present analysis with respect to the results presented in
Ref~\cite{marsella}, which include the dilepton channel as well.
This improvement is mainly due to:
\begin{itemize}
\item[(i)] The better sensitivity of the observables used. In the case of
$\gl$ and $\vr$ the improvement is moderate, with limits
about 1.13 times smaller. For $\gr$ the improvement is more significant,
by a factor of 1.34.
\item[(ii)] The combination of $\rhpm$ and $\Apm$.
\item[(iii)] The different statistical analysis used. For $\vr$ and
$\gl$, the Monte Carlo method used to obtain the true 68.3\% CL intervals
also reduces their size by 20\%--30\%, as explained above.
\end{itemize}
Finally, with the same procedure we obtain the 68.3\% CL confidence regions
on the anomalous couplings, presented in Fig.~\ref{fig:reg}. The boundary of the
regions has been chosen as a contour of constant $\chi^2$. In case that the
p.d.f. of $\vr$ and $\gl$ were Gaussian, the boundaries would be ellipses
corresponding to $\chi^2 = 2.30$ (see for instance Ref.~\cite{cowan}).
In our non-Gaussian case the $\chi^2$ for which the confidence regions have
68.3\% probability is determined numerically, and it is
approximately 1.83 for the $(\gl,\gr)$ plot and 1.85 for $(\vr,\gr)$.

\section{Conclusions}
\label{sec:6}

In this paper we have investigated the ATLAS sensitivity to non-standard $Wtb$
couplings. We have considered several observables: the helicity fractions
$\Fi$, helicity ratios $\rhpm$ and angular asymmetries $\afb$, $\Apm$. Although
these observables can be defined and measured for any top production process
with decay $t \to Wb \to \ell \nu b$, we have concentrated on top pair
production at LHC with semileptonic decay, with a large cross section and in
which the reconstruction of the final state is relatively easy.

Due to the excellent statistics available at LHC,
the precision reached is determined by systematic
uncertainties.
We have performed a very detailed study of the latter,
both theoretical ones and from the experimental reconstruction.
It has been found
that, although the observables considered are theoretically equivalent (as
noted in section~\ref{sec:2}), the systematic uncertainties in the measurement
of some of them, namely
$\rhp$ and $\Ap$, are smaller.
Since these observables also depend more strongly
on anomalous couplings, their measurement provides a more
sensitive probe for anomalous $Wtb$ couplings than helicity fractions. 
Moreover, when the four measurements of $\rhpm$ and $\Apm$
are combined, the sensitivity is further enhanced,
reaching the 5.5\%, 2.5\%
and 1.4\% level for $\vr$, $\gl$ and $\gr$ in Eq.~(\ref{ec:1}), respectively.
This is an important achievement for a hadronic machine.
Combining this measurement in $t \bar t$ semileptonic decays
with the dilepton decay channel $t \bar t \to \ell^+ \nu b \ell^{'-} \nu \bar b$
and single top production will improve (to what extent is yet to be
determined) these limits.

Although providing probably the strongest limits, the observables 
studied in this paper are not sufficient to fully constrain anomalous 
$Wtb$ couplings in a model-independent way. For nonzero $\vr$ and $\gl$, 
even of order $O(0.4)$, there are fine-tuned combinations for which 
their effects on helicity fractions and related observables almost 
cancel. Setting simultaneous limits on them requires additional 
observables with a different functional dependence on the $Wtb$ 
couplings. For example, in the dilepton channel two spin asymmetries 
$\all$ and $\allt$ involving the two leptons are found to be sensitive 
to $\vr$ but rather independent of $\gl$ \cite{paper1}. 
Ref.~\cite{marsella} has shown that these asymmetries can be measured 
with a good precision, 7\% and 5\%, respectively, and their study seems 
very promising. Spin asymmetries involving $b$ quarks like $\alb$ and 
$\albt$ exhibit a stronger dependence on anomalous couplings and when 
the appropriate detailed simulations are in place they will be studied. 
In addition single top production, involving $Wtb$ interactions in the 
production and the decay of the top quark will be studied, since it can 
provide complementary information about non-standard couplings through 
the cross sections for the different final states $tj$, $\bar tj$, $t 
\bar b$ and $\bar t b$, and spin asymmetries.

\vspace{1cm}
\noindent
{\Large \bf Acknowledgements}

\vspace{0.4cm} \noindent
This work has been performed within the ATLAS Collaboration, and we thank
collaboration members for helpful discussions.
We have made use of the physics analysis framework and tools which are the
result of collaboration-wide efforts. 
The work of J.A.A.-S. has been supported by a MEC Ramon y Cajal contract
and project FPA2006-05294, by
Junta de Andaluc{\'\i}a projects FQM 101 and FQM 437
and by the European Community's Marie-Curie Research Training
Network under contract MRTN-CT-2006-035505 ``Tools and Precision
Calculations for Physics Discoveries at Colliders''.
The work of J.C., N.C. (grant SFRH/BD/13936/2003), A.O. and F.V.
(grant SFRH/BD/18762/2004) has been supported by
Funda\c{c}\~ao para a Ci\^encia e a Tecnologia.

\end{document}